\documentclass[floatfix,
 reprint,
 superscriptaddress,
 amsmath,amssymb,
 aps,prl
]{revtex4-1}

\usepackage{mathrsfs} 
\usepackage{float,multirow}
\usepackage{mathrsfs}
\usepackage{upgreek}
\usepackage{xcolor}
\usepackage{mathtools}
\usepackage{amsmath,bm,morefloats}  
\usepackage{amsfonts} 
\usepackage{graphicx} 
\usepackage{CJK}

\usepackage{caption, subcaption}
\usepackage[multiple]{footmisc}
\newcommand{\mycomment}[1]{}
\begin{document}

\begin{CJK*}{UTF8}{}
 \CJKfamily{mj}

\title{Noise-Modified, Paraxial  Maxwell-Bloch Equations for X-ray Amplified Spontaneous Emission 
}

\author{Jeong-Wan Park}
\email{jeongwan.park@anl.gov}
\affiliation{Argonne National Laboratory, Lemont, IL 60439, USA}

\author{Kwang-Je Kim}
\affiliation{Argonne National Laboratory, Lemont, IL 60439, USA}
\affiliation{University of Chicago, Chicago, IL 60637, USA}
\author{Ryan Lindberg}
\affiliation{Argonne National Laboratory, Lemont, IL 60439, USA}

\date{\today}

\begin{abstract}
We present a Hamiltonian-based, 3D theory in paraxial approximation for X-ray amplified spontaneous emission (or superfluorescence) pumped by X-ray free-electron laser. The seed field is included. The ensemble-averaged Heisenberg equations become Maxwell-Bloch equations if factorization of operator products is assumed and are adequate when the stimulated emission is dominant. The spontaneous emission is accounted for by adding a random noise term to the atomic coherence, the magnitude of which is uniquely determined from the fact that the expectation value of the normally-ordered product electric field operators associated with an atom does not factorize. The modified Maxwell-Bloch equation we developed reproduces the results of the previous 1D theory based on correlation functions.

\end{abstract}

\maketitle 
\end{CJK*}

Successful observations of transient gain in atomic media via inner-shell photo-ionization \cite{nina,po2,po3,Doyle} and pumped by powerful X-ray free electron lasers \cite{emma, decking, ishikawa} have shown that X-ray lasers may be possible using the physical process variously referred to as superradiance~\cite{dicke}, superfluorescence~\cite{i2,i3}, or amplified spontaneous emission (ASE) \cite{laser}. The coherence and stability of an ASE-based X-ray laser can be  improved in an X-ray laser oscillator (XLO)  \cite{halavanau} by employing an X-ray cavity  as suggested for the X-ray FEL oscillator (XFELO) \cite{xfelo,xrafel}. 

ASE in the optical regime has been studied using Maxwell-Bloch equations (MBEs) that incorporate spontaneous emission in various ad hoc manners--see \cite{gross} for a review. They have been used also in the X-ray regime  \cite{po3,ex1,ns1,ex2,ns5,ns4,ns6,ex3,ns7,ex4}. 
However, since the time-scale for emission is comparable to that of the loss and decay processes at X-ray wavelengths, a more accurate treatment of spontaneous emission is needed.

In \cite{benediktovitch}, a 1D quantum theory for ASE was developed in which the 3D effects are parameterized by a small solid angle characterizing the pencil-shaped interaction region. Starting from the Heisenberg equations of motions for the field and atomic operators, a closed set of equations for the two-point correlation functions were derived in the absence of a seed field by assuming that the expectation value of products of operators from different atoms could be factorized. The theory incorporates spontaneous as well as stimulated emission.  

This letter presents a Hamiltonian-based, 3D theory of ASE including the seed field.  By employing the paraxial approximation, the Heisenberg equations of motion are  reduced to a form similar in structure to the 1D equations in \cite{benediktovitch}. The ensemble-averaged operator equations then are reduced to the MBEs \cite{gross} by enforcing factorization of operator products. However, the MBEs so obtained cannot be complete since products of two operators from the same atom do not factorize. This problem is  resolved by adding a random phase term to the atomic coherence, the magnitude of which is determined by considering the average of the photo-current operator. The modified MBEs include spontaneous emission and result in two point correlation functions that agree with those of \cite{benediktovitch} in 1D limit. The modified MBEs also predict the correct time behavior of the spontaneous emission, i.e., the exponential decay.

Let \(\Omega\) be the angular frequency corresponding to the energy difference between the ground \(|g\rangle\) and excited \(|e\rangle\) states of each atom, while \(a_{\mathbf{k}, s}^{\dagger}\) and \(a_{\mathbf{k}, s}\) are the creation and annihilation operators, respectively, of photons with wave vector \(\mathbf{k}\), frequency \(\omega_{\mathbf{k}}=ck\), and polarization index \(s\). We introduce  atomic operators
\begin{equation}\label{sigmas}
\sigma_{+}=|e\rangle\langle g |, \quad \sigma_{-}=|g\rangle\langle e |, \quad \sigma_{z}=|e\rangle\langle e|-| g\rangle\langle g| ,
\end{equation}
which satisfy the commutation relations \(\left[\sigma_+^{(a)},\sigma_-^{(b)}\right]=\delta_{ab}\sigma_z\), 
\(\left[\sigma_z^{(a)},\sigma_{\pm}^{(b)}\right]=\pm 2\delta_{ab}\sigma_{\pm}\).
We assume the electric-dipole transition matrix element to be in the \(x\)-direction and is real with  magnitude \(\mu\). The positive-electric field operator in the \(x\)-direction at the location of atom \(a\) and its conjugate  are  
\begin{equation}\label{Efield}
E_{+}^{(a)}(t)=\sum_{\mathbf{k},s} \cos{\vartheta_{\mathbf{k},s}} \mathscr{E}_{\mathbf{k}} e^{i \mathbf{k} \cdot \mathbf{r}_{a}}  a_{\mathbf{k}, s}(t), ~ E_{-}^{(a)}(t)={E^{(a)\dagger}_{+}(t)},
\end{equation}
where \(\vartheta_{\mathbf{k},s}\) is the angle between the \(x\) direction and the \(s-\)polarization, \(\mathscr{E}_{\mathbf{k}}= \sqrt{\frac{\hbar\omega_{\mathbf{k}}}{2 \varepsilon_{0} V }}\), \(V\) is the quantization volume, and \(\mathbf{r}_{a}\)  the position of atom \(a\). The  notations of \cite{scully} and the SI units are used throughout this Letter. Solving the Heisenberg equation of motion for \(a_{\mathbf{k}, s}\) and inserting to Eq. \eqref{Efield} results in
\begin{equation}\label{E(t)}
E_{+}^{(a)}(t)=\frac{i\hbar}{2\mu} \Gamma_{\mathrm{sp}} \sigma_{-}^{(a)}(t)+E_{+}^{\prime(a)}(t).  
\end{equation}
Here \(\Gamma_{\mathrm{sp}}=\frac{|\boldsymbol{\mu}|^2}{3\pi \varepsilon_0 \hbar} \left(\frac{\Omega}{c}\right)^3 \) is the spontaneous emission rate and
\begin{widetext}
\begin{equation}\label{Eprime}
    E_{+}^{\prime(a)}(t)=\sum_{\mathbf{k},s} \cos{\vartheta_{\mathbf{k},s}}\left[\mathscr{E}_{\mathbf{k}}  a_{\mathbf{k}, s}(0) e^{-i\omega_{\mathbf{k}}t+i\mathbf{k}\cdot \mathbf{r}_a} + i \frac{\omega_{\mathbf{k}}\mu}{2\varepsilon_0 V} \cos{\vartheta_{\mathbf{k},s}}\int_0^{t} d t^{\prime} \sum_{b \neq a } e^{-i\omega_{\mathbf{k}}\left(t-t^{\prime}\right)+i\mathbf{k}\cdot \left(\mathbf{r}_a -\mathbf{r}_{b}\right)} \sigma_{-}^{(b)}(t^{\prime})\right] .
\end{equation}
\end{widetext}
The first term in Eq.~\eqref{E(t)} arises from  the omitted term \(b=a\) in Eq.~\eqref{Eprime} evaluated using the Weisskopf-Wigner approximation \cite{scully}.
The Heisenberg equations of motion for the \(\sigma\) operators are
\begin{align}
\frac{d \sigma_{-}^{(a)}}{d t}
=&-\left(i \Omega +\frac{\Gamma_{\mathrm{sp}}}{2}\right) \sigma_{-}^{(a)}(t)-\frac{i\mu}{\hbar}\sigma_{z}^{(a)}(t)E_{+}^{\prime( a)}(t), \label{dsig-dt} \\
\begin{split}
\frac{d \sigma_{z}^{(a)}}{d t}=&
-2\Gamma_{\mathrm{sp}}\sigma_{+}^{(a)}(t) \sigma_{-}^{(a)}(t) \\&+\frac{2i\mu}{\hbar} \left[\sigma_{+}^{(a)}(t) E_{+}^{\prime( a)}(t)- E_{-}^{\prime( a)}(t)\sigma_{-}^{(a)}(t)\right].  \label{dsigzdt}
\end{split}
\end{align}

We consider the case where the atomic system is pumped by XFEL radiation of narrow divergence which thereby defines a ``thin-pencil'' interaction volume of length $L$ and radius $R$, with $R\ll L$.  We define \(z\) to be the coordinate along the pencil axis with $\mathbf{x}=(x,y)$ the transverse coordinates, so that the the position vector \(\mathbf{r}=(\mathbf{x}, z)\).  We then introduce  \(k_{\Omega}=\Omega/c\)  and \(\lambda_{\Omega}=2\pi /k_{\Omega}\) under the assumption that \(\lambda_{\Omega} \ll R\ll L\), and approximate the discrete sum over \(\mathbf{k}\) in  Eq.~\eqref{Eprime} using the integral \( \frac{V}{8 \pi^3}\int d^3 \mathbf{k} \).  We will evaluate the latter in terms of the magnitude \(k=\omega_{\mathbf{k}}/c\) and the transverse angles \(\boldsymbol{\phi}=\left(\phi_x, \phi_y\right)\).  The old and new variables are related by  \(\mathbf{k}=k\left (\boldsymbol{\phi},1-\boldsymbol{\phi}^2/2\right)\) in the \textit{paraxial} approximation for which \(|\boldsymbol{\phi}| \ll 1\).  The exponential factor in the second, inhomogeneous term of \eqref{Eprime} becomes
 \begin{equation}\label{exponential}
   e^{i\Phi+i\Delta \Phi+ik \left[(z_a-z_b)-c\left(t-t^{\prime}\right)\right]},
 \end{equation}
where
\begin{align}\label{Phi}
 \Phi &= k_{\Omega}\left[\boldsymbol{\phi} \cdot \left(\mathbf{x}_a-\mathbf{x}_b\right)-\frac{\boldsymbol{\phi}^2}{2} \left(z_a-z_b\right)\right],   \\ 
\label{DelPhi}
\Delta \Phi &= \Delta k \left[\boldsymbol{\phi} \cdot \left(\mathbf{x}_a-\mathbf{x}_b\right)-\frac{\boldsymbol{\phi}^2}{2}\left(z_a-z_b\right)\right] ,  
\end{align}
and \(\Delta k=k-k_{\Omega}\).

Since the interactions occur near the atomic transition frequency \(\Omega\), the main contribution to the \(k\)-integral comes from  the region \(\Delta k \ll k_{\Omega}\). We therefore neglect \(\Delta\Phi\) and replace \(k\rightarrow k_{\Omega}\) in all other places in the integrand.
The main contribution of the \(\boldsymbol{\phi}\)-integral would come from the region where \(|\Phi| \lesssim \pi\). In view of Eq.~\eqref{Phi}  this corresponds to \(|\boldsymbol{\phi}| \lesssim \lambda_{\Omega}/2 R\) and \(\quad |\boldsymbol{\phi}| \lesssim\sqrt{\lambda_{\Omega}/L }\); from geometry we also have
\(\left|\boldsymbol{\phi} \right| \lesssim R/L\).

For X-ray ASE under consideration here, all the three limits of the angle are much less than unity, validating the paraxial approximation. It then follows that
 \begin{equation}\label{kintegral}
   \int d^3 \mathbf{k}  \approx \int d^2\boldsymbol{\phi}\int_0^{\infty} k^2 dk  \approx  \int d^2\boldsymbol{\phi}\int_{-\infty}^{\infty} k_{\Omega}^2 d\Delta k   .
 \end{equation}
Here, we have neglected the wave propagating opposite to the pump wave, namely, along the positive \(z\)-direction.

The polarization vectors for \(s=1, 2\) can be taken approximately parallel to \(x,y\), respectively. Thus, \(\vartheta_{\mathbf{k},1} \approx 0\) and \(\vartheta_{\mathbf{k},2} \approx \pi/2\) and the \(\Delta k-\)integral becomes \(\delta\left(z_a-z_b-c\left(t-t^{\prime}\right)\right)\), rendering the \(t^{\prime}\)-integral trivial.  Finally, we follow \cite{benediktovitch} and introduce the retarded time \(\tau=t-z_b/c\) associated with the atomic location \(b\), so that the slowly varying part \(\tilde{E}_{+}^{(a)}(\tau) = e^{i\Omega \tau}E_+^{\prime (a)}\) becomes
\begin{equation}\label{Eslow}
\begin{aligned}
    \tilde{E}_{+}^{(a)}(\tau) &=
    \tilde{E}_{+,in}^{(a)}(\tau)
    +\frac{3i\hbar \Gamma_\mathrm{sp}}{
    8 \pi \mu}\sum_{z_b < z_a}    \mathcal{G}(\mathbf{r}_a-\mathbf{r}_b)\tilde{\sigma}_{-}^{(b)}(\tau) .
    \end{aligned}
\end{equation}
Here, 
\begin{equation}\label{paragreen}
\mathcal{G}(\mathbf{r}) =
\int_{-\infty}^{\infty}d^2 \boldsymbol{\phi} e^{i\frac{\Omega}{c}\left(\boldsymbol{\phi}\cdot \mathbf x-\frac{\boldsymbol{\phi}^2 }{2} z\right)}=\frac{\lambda_{\Omega}}{i z}e^{i\frac{\pi \mathbf{x}^2}{\lambda_{\Omega} z}}
\end{equation} 
is the paraxial approximation of the far-field Green function \cite{loudon}.  The operator \(\tilde{E}_{+,in}^{(a)}(\tau)\) is the envelope of the incoming paraxial seed field evaluated at the atomic location \(a\), and corresponds to the term containing the operator \(a_{\mathbf{k}, s}(0)\) in Eq.~\eqref{Eprime}. The seed field is essential when studying multi-pass lasers such as the XLO.

We can reduce Eq.~\eqref{Eslow} to the 1D equations used in \cite{benediktovitch} by averaging \(\mathcal{G}(\mathbf{r}_a-\mathbf{r}_b)\) over the transverse positions of atoms \(a\) and \(b\).  We obtain this limit by setting \(\mathbf{r}_a=(\mathbf{x},L)\), \(\mathbf{r}_b=(\mathbf y, L-\Delta z)\), and assuming that the transverse density profile of the pumped region is Gaussian with the RMS size \(\sigma_x=R/\sqrt{2}\):
\begin{align}
\mathcal{G}_{av}(\Delta z)&=\frac{1}{\left(\pi R^2\right)^2}\int d\mathbf{x}d\mathbf{y} e^{-\frac{\mathbf{x}^2+\mathbf{y}^2}{R^2}}\mathcal{G}\left(\mathbf{x}-\mathbf{y},\Delta z\right) \nonumber \\
&=\frac{\lambda_\Omega^2}{ 2\pi R^2+i \Delta z\lambda_\Omega}. \label{Gav}
\end{align}
In \cite{benediktovitch}, the quantity \(2 \mathcal{G}_{av}\) was chosen to be the constant solid angle \(\Delta o\) \cite{footnote1}. We will find that the important region is \(\Delta z\ll \Delta z_{ch}\equiv 2\pi R^2/\lambda_\Omega\), leading to the identification \(\Delta o \approx \Delta o_{3d}= \lambda_\Omega^2/\pi R^2\). Geometry also suggests the possible constant \(\Delta o=\Delta o_g\equiv \pi R^2/L^2\) \cite{benediktovitch}.  We will compare our 3D theory with 1D approximations using these constants later.

The slowly-varying version of Eqs.~\eqref{dsig-dt} and \eqref{dsigzdt} are
\begin{align}
\frac{d \tilde{\sigma}_{-}^{(a)}}{d\tau}
=&-\frac{\Gamma_{\mathrm{sp}}}{2} \tilde{\sigma}_{-}^{(a)}(\tau) -\frac{i\mu}{\hbar}\sigma_{z}^{(a)}(\tau)\tilde{E}_{+}^{( a)}(\tau), \label{dsig-dtslow} \\
\begin{split}\label{dsigzdtslow}
\frac{d \sigma_{z}^{(a)}}{d \tau}=&
-2\Gamma_{\mathrm{sp}}\tilde{\sigma}_{+}^{(a)}(\tau) \tilde{\sigma}_{-}^{(a)}(\tau) \\&+\frac{2i\mu}{\hbar} \left[\tilde{\sigma}_{+}^{(a)}(\tau) \tilde{E}_{+}^{( a)}(\tau)- \tilde{E}_{-}^{(a)}(\tau)\tilde{\sigma}_{-}^{(a)}(\tau)\right].
\end{split}
\end{align}
Note \(\sigma_z\) is a slowly varying quantity and does not need the tilde notation.
Equations \eqref{Eslow}, \eqref{dsig-dtslow}, and \eqref{dsigzdtslow} are the 3D generalization of the corresponding 1D equations derived in \cite{benediktovitch}. Due to the particular way the paraxial approximation was applied, the dependence on the longitudinal variable \(\tau\) is factored out which results in the 3D equations having the same structure as those in 1D.  

We now consider the ensemble average  of operators indicated by angular brackets. We begin with the electric field Eq.~\eqref{Eslow}
\begin{align}
&\mathcal{E}_{+}^{(a)}(\tau)\equiv  \left\langle \tilde{E}_{+}^{(a)}(\tau)\right\rangle  \nonumber \\
&= \mathcal{E}_{+,in}^{(a)}(\tau)+\frac{3i\hbar\Gamma_\mathrm{sp}}{
8 \pi\mu }\sum_{z_b < z_a} \mathcal{G}(\mathbf{r}_a-\mathbf{r}_b)\rho_{eg}^{(b)}(\tau) ,  \label{Eav}
\end{align}
where \( \mathcal{E}_{+,in}^{(a)}(\tau)=\langle E_{+,in}^{(a)} (\tau)\rangle\) gives the spatio-temporal profile of the incoming seed field, and
\begin{equation}
\rho_{eg}^{(a)}=\left\langle \tilde{\sigma}_-^{(a)}\right\rangle=\mathrm{Tr}\left(\tilde{\sigma}_-^{(a)}\rho^{(a)}\right)
\end{equation}
with \(\rho^{(a)}_{ij}\) the \((i,j)\) element of the atomic density matrix \(\rho^{(a)}\). 
If we adopt a continuous description by replacing discrete atomic labels by their average positions \(\mathbf {r}\), the field \(\mathcal{E}_{+}(\mathbf{r}, \tau)\) obtained from Eq.(\ref{Eav}) satisfies the paraxial Maxwell equation in differential form given by \cite{gross,kai}.

In taking the average of Eqs. \eqref{dsig-dtslow} and \eqref{dsigzdtslow}, we make a crucial assumption necessary to obtain a finite, closed system of equations: we assume that averages of products of operators factorize. Thus, for \(a \neq b\)
\begin{align}
\left\langle\sigma_{z}^{(a)}(\tau) \tilde{\sigma}_{-}^{(b)}(\tau) \right\rangle & \approx \left\langle\sigma_{z}^{(a)}(\tau)\right\rangle\left\langle\tilde{\sigma}_{-}^{(b)}(\tau)\right\rangle,
\label{bbgky} \\
    \left\langle \tilde{\sigma}_{+}^{(a)}(\tau) \tilde{\sigma}_{-}^{(b)}(\tau)\right\rangle &\approx \left\langle\tilde{\sigma}_{+}^{(a)}(\tau)\right\rangle \left\langle\tilde{\sigma}_{-}^{(b)}(\tau)\right\rangle,  \label{bbgky2}
\end{align}
and we thereby obtain
\begin{equation}\label{drhoge}
\frac{d \rho_{ge}^{(a)}}{d \tau}=-\frac{\Gamma^{(a)}}{2} \rho_{ge}^{(a)}( \tau)-\frac{i \mu}{\hbar} \rho_{\mathrm{inv}}^{(a)}( \tau) \mathcal{E}_{+}^{(a)}(\tau).
\end{equation}
Here, \(\rho_{\mathrm{inv}}^{(a)} \equiv \left\langle\sigma_{z}^{(a)}(\tau)\right\rangle=\rho_{e e}^{(a)}(\tau)-\rho_{gg}^{(a)}(\tau)\) and \(\mathcal{E}_{+}^{(a)}\) is given by \eqref{Eav}. 
Similarly, it follows from Eq.~\eqref{dsigzdtslow} that
\begin{align}
\begin{split}
\frac{d \rho_{e e}^{(a)}}{d \tau}=&r_e^{(a)}(\tau)-\Gamma_{ee}^{(a)}(\tau) \rho_{e e}^{(a)}( \tau) \\
&+\frac{ i \mu}{\hbar}\left[ \mathcal{E}_{+}^{(a)}(\tau) \rho_{g e}^{(a)}( \tau)-\mathcal{E}_{-}^{(a)}(\tau) \rho_{ eg}^{(a)}(\tau) \right], \label{drhoee}
\end{split} \\
\begin{split}
\frac{d \rho_{g g}^{(a)}}{d \tau}=&r_g^{(a)}(\tau)+\left(\Gamma_{\mathrm{sp}}+\gamma_n\right) \rho_{e e}^{(a)}( \tau)-\gamma_g^{(a)}(\tau)\rho_{gg}^{(a)}(\tau) 
\\&-\frac{ i \mu}{\hbar}\left[ \mathcal{E}_{+}^{(a)}( \tau) \rho_{g e}^{(a)}( \tau)-\mathcal{E}_{-}^{(a)}( \tau) \rho_{ eg}^{(a)}( \tau) \right].  \label{drhogg}
\end{split}
\end{align}
In the above, we have introduced additional incoherent processes using the Lindblad superoperator method following \cite{benediktovitch}, and introduced various rates as follows:
\begin{equation}\label{gamee}
\begin{split}
\Gamma^{(a)}(\tau)&\equiv\Gamma_{ee}^{(a)}(\tau)+\gamma_g^{(a)}(\tau)+q^{(a)}(\tau),  \\
    \Gamma_{ee}^{(a)}(\tau)&\equiv\Gamma_\mathrm{sp}+\gamma_e^{(a)}(\tau)+\gamma_n.
    \end{split}
\end{equation}
Here \(\gamma_n\) is the non-radiative decay rate,  \(q\) is the rate for atoms' decoherence, \(r_e\) and \(r_g\) are pumping rates for the excited state and ground state respectively, and \(\gamma_e\) and \(\gamma_g\) are depletion rates for the excited and ground states, respectively, that are mainly due to auger decay.

Equations \eqref{Eav}, \eqref{drhoge}, \eqref{drhoee}, and \eqref{drhogg} form  a closed set of MBEs. Gain and saturation in a pencil shaped medium can be computed from these equations; see \cite{gross} for a review. 
However, the MBEs cannot serve as a genuine model for ASE since they do not capture spontaneous emission.  To remedy this, we begin by noting that the positive frequency part of the electric field originating from atom \(a\) is, in view of the discussion leading to Eq.~\eqref{Eslow}, proportional to \(\tilde{E}_{+} (\tau) \propto \tilde{\sigma}_{-}^{(a)}(\tau) \).  The intensity of the associated photo-current is \cite{loudon}
\begin{align}
&\left\langle\tilde{E}_{-}(\tau)\tilde{E}_{+}(\tau)\right\rangle \propto \left\langle \tilde{\sigma}_{+}^{(a)}(\tau)\tilde{\sigma}_{-}^{(a)}(\tau)\right\rangle \nonumber \\ 
&= \rho_{ee}^{(a)}(\tau) = \left|\rho_{ge}^{(a)}(\tau)\right|^2+\left(\rho_{ee}^{(a)}(\tau) -\left|\rho_{ge}^{(a)}(\tau)\right|^2\right).  \label{photocurrent}
\end{align}
The term \(\rho_{ge}\rho_{eg}=\left|\rho_{eg}\right|^2\) is what we would get if we assume the factorization used in Eq.~\eqref{bbgky2}.  This, however, is not valid since \(a=b\), and the term in parentheses is the necessary correction that we identify with spontaneous emission. We do this by adding a random noise amplitude \(\xi^{(a)} \) to \(\rho_{eg}^{(a)}\) as follows:
\begin{equation}\label{noise12}
\rho_{ge}^{(a)}(\tau) \rightarrow \hat{\rho}_{ge}^{(a)}(\tau) \equiv \rho_{ge}^{(a)}(\tau)+\xi^{(a)}(\tau).
\end{equation}
The noise amplitude \(\xi^{(a)}(\tau)\) has the properties
\begin{equation}\label{xi1}\begin{aligned}
   \left \langle \xi^{(a)}(\tau) \right\rangle_{en}&=0, 
   \\
   \left \langle \xi^{(a)}(\tau_1) \xi^{*(b)}(\tau_2) \right\rangle_{en} &= \left(\rho_{ee}^{(a)}(\tau_1)-\left|\rho_{ge}^{(a)}(\tau_1)\right|^2\right) \delta_{a b}\delta_{\tau_1 \tau_2},
\end{aligned}\end{equation}
where \( \left \langle ... \right\rangle_{en}\) indicates taking the ensemble average.  The modified MBEs in which \(\rho_{ge}^{(a)}(\tau) \) is replaced by \(\hat{\rho}_{ge}^{(a)}(\tau)\) correctly account for spontaneous emission and thus can be used to model X-ray ASE. 

The contribution of spontaneous emission to the dynamics of ASE was found in \cite{benediktovitch} by computing the derivative of the two point correlation function \(d \big\langle \sigma_+^{(a)}\sigma_-^{(b)}\big\rangle/d\tau\)  for \(a \neq b\). Using the Heisenberg equations, this quantity can be  expressed as a sum of the average of operator products. Imposing the factorization of Eqs.~\eqref{bbgky} and \eqref{bbgky2}, even for a term  proportional to \(\left\langle \sigma_+^{(c)}\sigma_-^{(c)} \right\rangle = \rho_{ee}^{(c)}\) that cannot be factorized in principle, results in the MBEs (namely, Eq.~\eqref{drhoge} and its conjugate). Ref.~\cite{benediktovitch} identified this term as what accounts for spontaneous emission, and we have shown that their result can be obtained from the usual MBEs by replacing \(\rho_{ge}^{(c)}(\tau) \) with \(\hat{\rho}_{ge}^{(c)}(\tau)\).

Since the noise amplitude changes abruptly from one instant to the next, care must be exercised in numerically integrating the modified MBE equations.  Introducing index \((n)\) for discrete time steps \(\tau_n=n\Delta\tau\), we proceed as follows: Let  \(\rho_{ee(n)}^{(a)}\), \(\rho_{gg(n)}^{(a)}\), and \(\rho_{eg(n)}^{(a)}\) be given. Then we write
\begin{equation}\label{rhohatn}
\hat{\rho}_{ge(n)}^{(a)}=\rho_{ge(n)}^{(a)}+\xi^{(a)}_{(n)},
\end{equation}
where
\begin{equation}\label{xi2}
\xi^{(a)}_{(n)}=\sqrt{\rho_{ee(n)}^{(a)}-\left|\rho_{ge(n)}^{(a)}\right|^2}e^{i\Phi^{(a)}_{(n)}}.
\end{equation}
Here \(\Phi_{(n)}^{(a)}\) is the random phase with the property:
\begin{equation}\label{rphase}
 \left\langle e^{i\left(\Phi^{(a)}_{(n)}-\Phi^{(b)}_{(m)}\right)}\right\rangle_{en}=\delta_{ab}\delta_{nm}. 
\end{equation}
The \(n\)th step is complete with
\begin{equation}\label{Ehatn}
\hat{\mathcal{E}}_{+(n)}^{(a)}= \mathcal{E}_{+,in(n)}^{(a)}+i\frac{3\hbar}{
8 \pi\mu }\Gamma_\mathrm{sp}\sum_{z_b < z_a} \mathcal{G}(\mathbf{r}_a-\mathbf{r}_b)\hat{\rho}_{eg(n)}^{(b)} .
\end{equation}
We are ready for the next time step beginning with the quantities
\begin{widetext}\begin{equation}\begin{alignedat}{2}\label{np1}
\rho_{ee(n+1)}^{(a)}&=\rho_{ee(n)}^{(a)}+\left[r_{e(n)}^{(a)}-\Gamma_{ee(n)}^{(a)} \rho_{e e(n)}^{(a)} +\frac{ i \mu}{\hbar}\left[ \hat{\mathcal{E}}_{+(n)}^{(a)} \hat{\rho}_{g e(n)}^{(a)}-\hat{\mathcal{E}}_{-(n)}^{(a)} \hat{\rho}_{ eg(n)}^{(a)} \right]\right]\Delta \tau,\\ 
\rho_{gg(n+1)}^{(a)}&=\rho_{gg(n)}^{(a)}+\left[r_{g(n)}^{(a)}+\left(\Gamma_{\mathrm{sp}}+\gamma_n\right) \rho_{e e(n)}^{(a)}-\gamma_{g(n)}^{(a)}\rho_{gg(n)}^{(a)} -\frac{ i \mu}{\hbar}\left[ \hat{\mathcal{E}}_{+(n)}^{(a)} \hat{\rho}_{g e(n)}^{(a)}-\hat{\mathcal{E}}_{-(n)}^{(a)} \hat{\rho}_{ eg(n)}^{(a)} \right]\right]\Delta \tau,\\
{\rho}_{ge(n+1)}^{(a)}&=\hat{\rho}_{ge(n)}^{(a)}+\left[-\frac{\Gamma^{(a)}_{(n)}}{2} \hat{\rho}_{g e(n)}^{(a)}+\frac{i \mu}{\hbar} \rho_{\mathrm{inv}(n)}^{(a)} \hat{\mathcal{E}}_{-(n)}^{(a)}\right]\Delta \tau , 
\end{alignedat}
\end{equation}
\end{widetext}
and computing \(\hat{\rho}_{ge(n+1)}^{(a)}\) and \(\hat{\mathcal{E}}_{+(n+1)}^{(a)}\) by substituting \(n\rightarrow n+1\) in Eqs. \eqref{rhohatn} and \eqref{Ehatn}.

To test the above recipe for numerical integration, we apply it to the case of spontaneous emission of a single atom \(a\). We assume no incoherent process other than the pumping to the excited state, i.e., \(\Gamma_{ee}=\Gamma_{\mathrm{sp}}, \gamma_n=\gamma_g=r_g=0\), where for simplicity we have suppressed the atomic label. Since no other atoms are present, the electric field \(\hat{\mathcal{E}}_-\) vanishes assuming no incoming field, and
\begin{equation}\label{spontrhoee}
\begin{aligned}
\left\langle \left| \hat{\rho}_{ge}(n) \right|^2 \right\rangle_{en}&=\rho_{ee}(n)\\
&=X^n \rho_{ee}(0)+\sum_{m=1}^{n}X^{n-m} r_{e}(m)\Delta\tau.
\end{aligned}
\end{equation}
Here \(X=1-\Gamma_{\mathrm{sp}}\Delta\tau\), so that for a small time step, \(\Delta \tau \ll \Gamma_{\mathrm{sp}}^{-1}\), we have \(X^m=e^{-\Gamma_{\mathrm{sp}}\tau_m}\).  
Both the atomic excitation and the intensity of the spontaneous emission decay exponentially with with the e-folding length of \(\Gamma_{\mathrm{sp}}^{-1}\). The excitation is replenished by \(r_{e}(m)\) due to pumping at \(\tau_m\) but it also decays exponentially.   By computing the far-field electric field \cite{loudon} and Poynting vector, one can check that the number of photons emitted is equal to the initial excitation \(\rho_{ee}(0)\) in the absence of pumping.

Spontaneous emission has been often modeled by adding a random phase term to the {\it time derivative} of \(\rho_{ge}\) \cite{ns4,ns10,ns11,benediktovitch2} rather than to \(\rho_{ge}\) itself  as we do in Eq.~\eqref{rhohatn}. 
In that case the early-time dynamics resembles Brownian motion and the intensity grows linearly in time. When combined with spontaneous emission decay the intensity evolution then behaves as \(\Gamma_\mathrm{sp} \tau e^{-\Gamma_\mathrm{sp} \tau}\) and exhibits an incorrect, time-delayed peak, as noted in \cite{benediktovitch}.

\begin{figure*}
       \begin{subfigure}[t]{0.33\linewidth}
       \centering
\includegraphics[width=\textwidth]{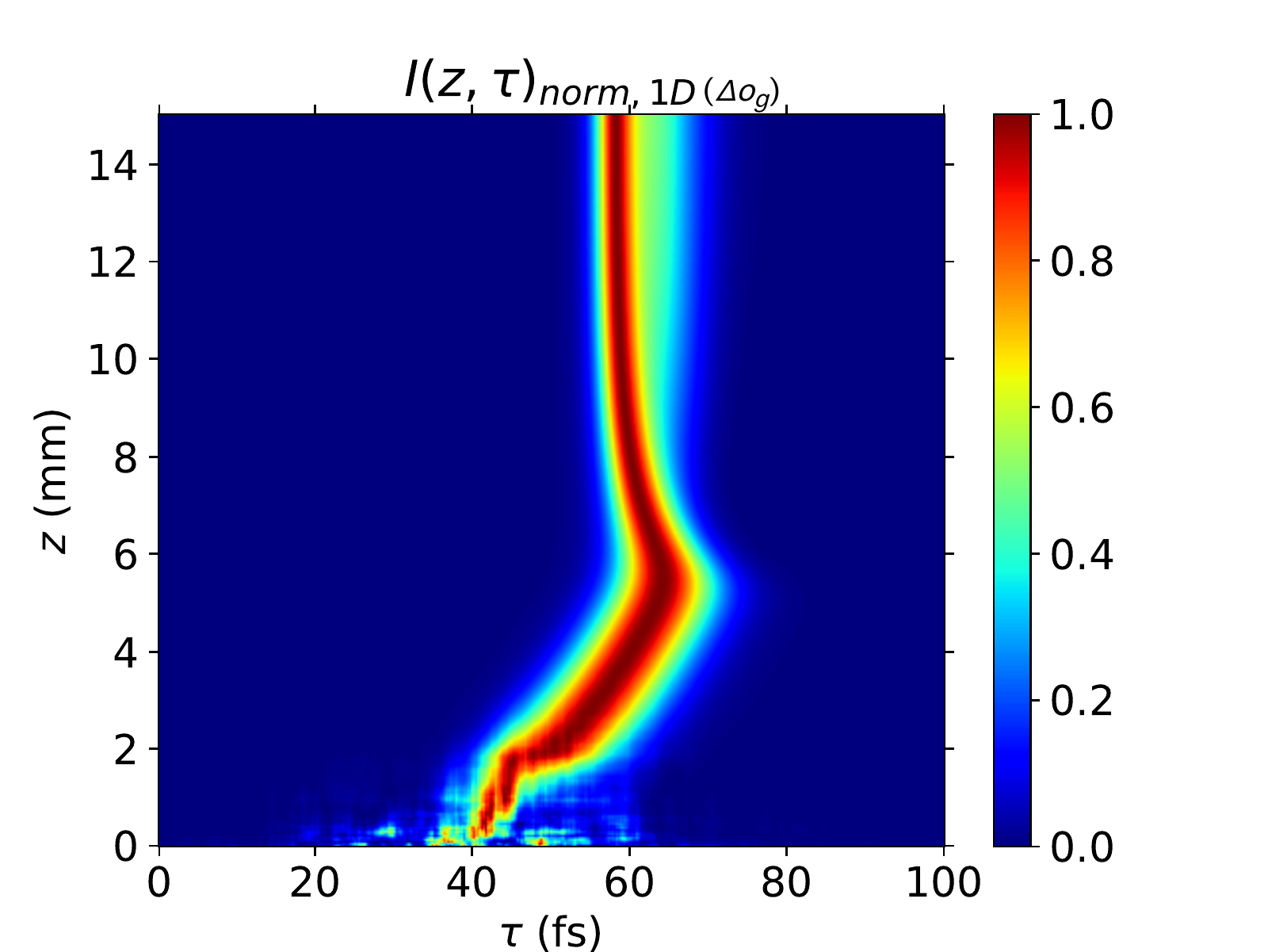}
         \caption{}
         \label{one}
     \end{subfigure}\hfill
         \begin{subfigure}[t]{0.33\linewidth}
         \centering
\includegraphics[width=\textwidth]{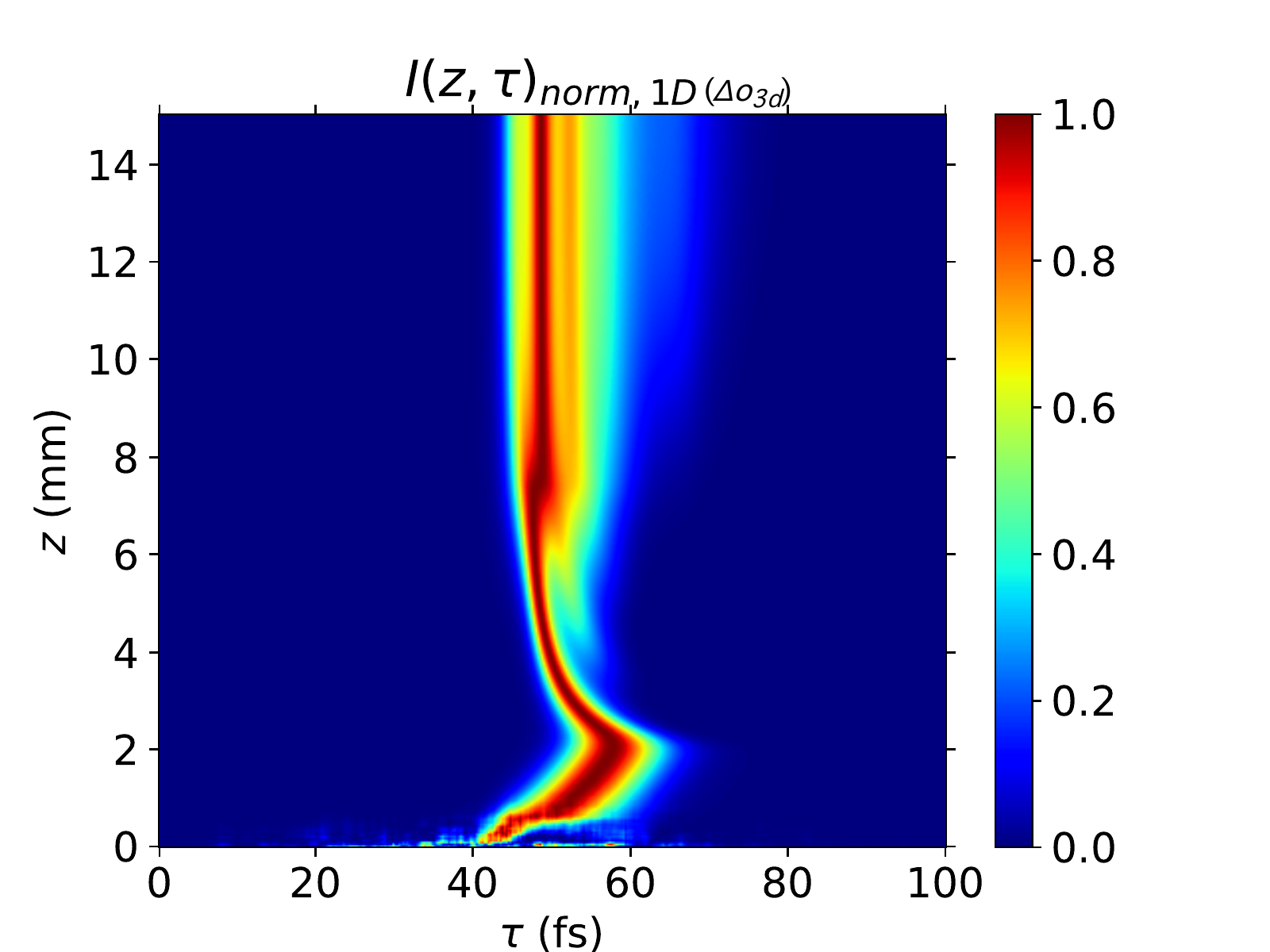}
         \caption{}
         \label{one_2}
     \end{subfigure}\hfill
         \begin{subfigure}[t]{0.33\linewidth}
\includegraphics[width=\textwidth]{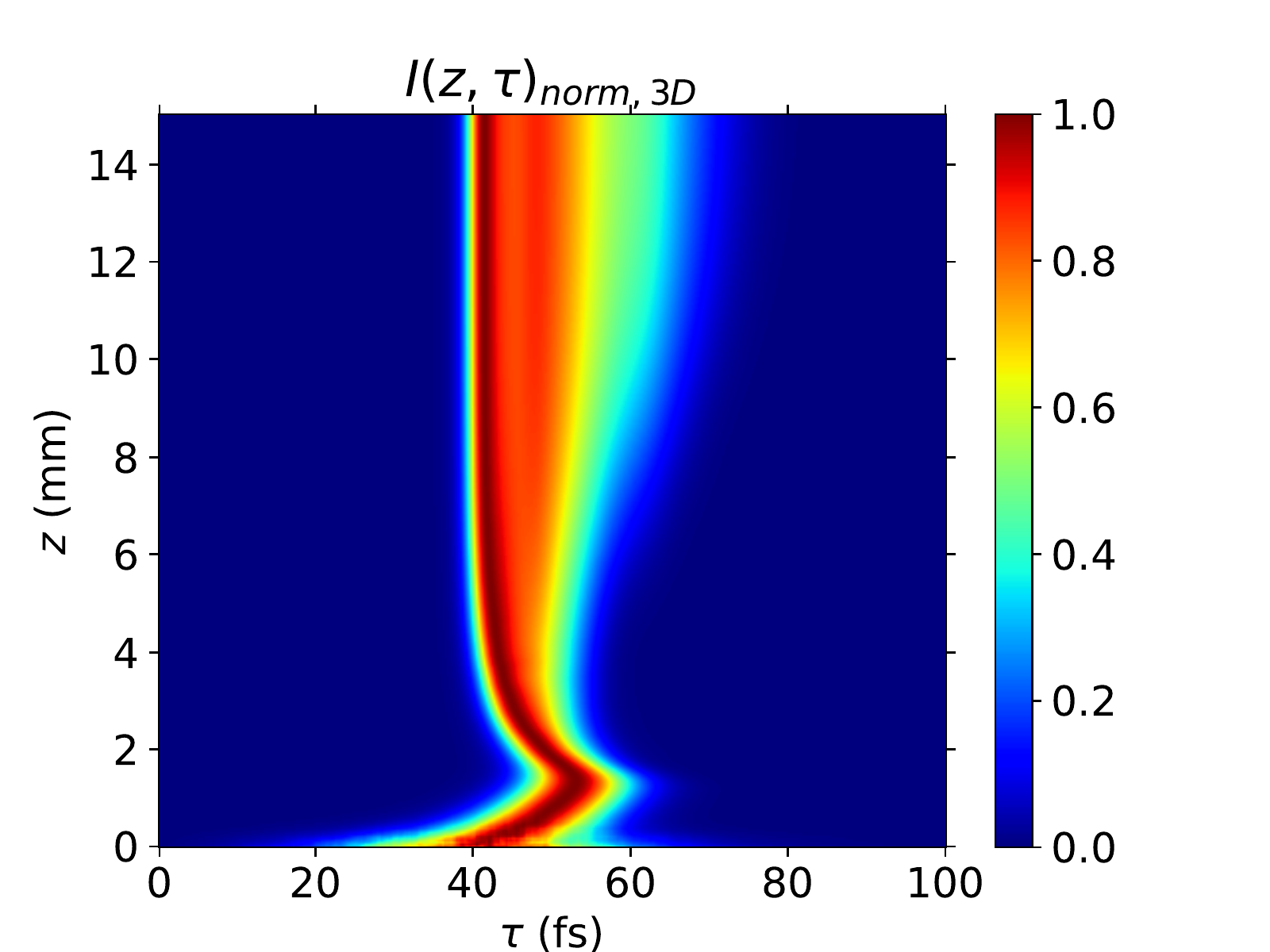}
         \caption{}
         \label{average}
        \end{subfigure}\hfill
     \caption{Radiation intensity profile from (a) modified 1D MBEs with \(\Delta o_g\), (b) modified 1D MBEs with \(\Delta o_{3d}\), and (c) modified 3D MBEs, normalized at each value of \(z\). }
        \label{converge}
    \end{figure*}

Further details of the simulation program and calculation of XLO performance both in low- and high-gain regimes will be presented in a later publication \cite{exp_paper}. Here we give some simulation results using parameters corresponding to the experiment described in \cite{nina}; \(\lambda_\Omega=1.46~\mathrm{nm}\), \(\Gamma_\mathrm{sp}^{-1}=160~\mathrm{fs}\), \(\gamma_e^{-1}=2.4~\mathrm{fs}\), \(R=2~\mu \mathrm{m}\), \(L=15~\mathrm{mm}\), atomic volume density \(n_v=1.6\times 10^{25}~\mathrm{m}^{-3}\), \(r_g=\gamma_g=q=\gamma_n=0\), photoionization cross section of pumping to the excited state \(\sigma_\mathrm{abs}=0.3~\mathrm{Mb}\), \(880~\mathrm{eV}\) XFEL pump pulse  containing \(N_{ph,pump}=2\times 10^{12}\) photons within \(40~\mathrm{fs}\) FWHM Gaussian temporal profile \cite{benediktovitch}. A flat-top transverse profile is used for the pump in our modified 3D MBEs, and \(\Delta o=\Delta o_g\equiv \pi R^2/L^2=0.56 \times 10^{-7}\) or   \(\Delta o = \Delta o_{3d}\equiv  \lambda_\Omega^2/\pi R^2=1.70 \times 10^{-7}\) for 1D simulation. For the parameters used in these simulation, \(\Delta z_{ch}\sim 17 ~\mathrm{mm}\).

\begin{figure}[h]
       \begin{subfigure}[t]{0.5\linewidth}
       \centering
\includegraphics[width=\textwidth]{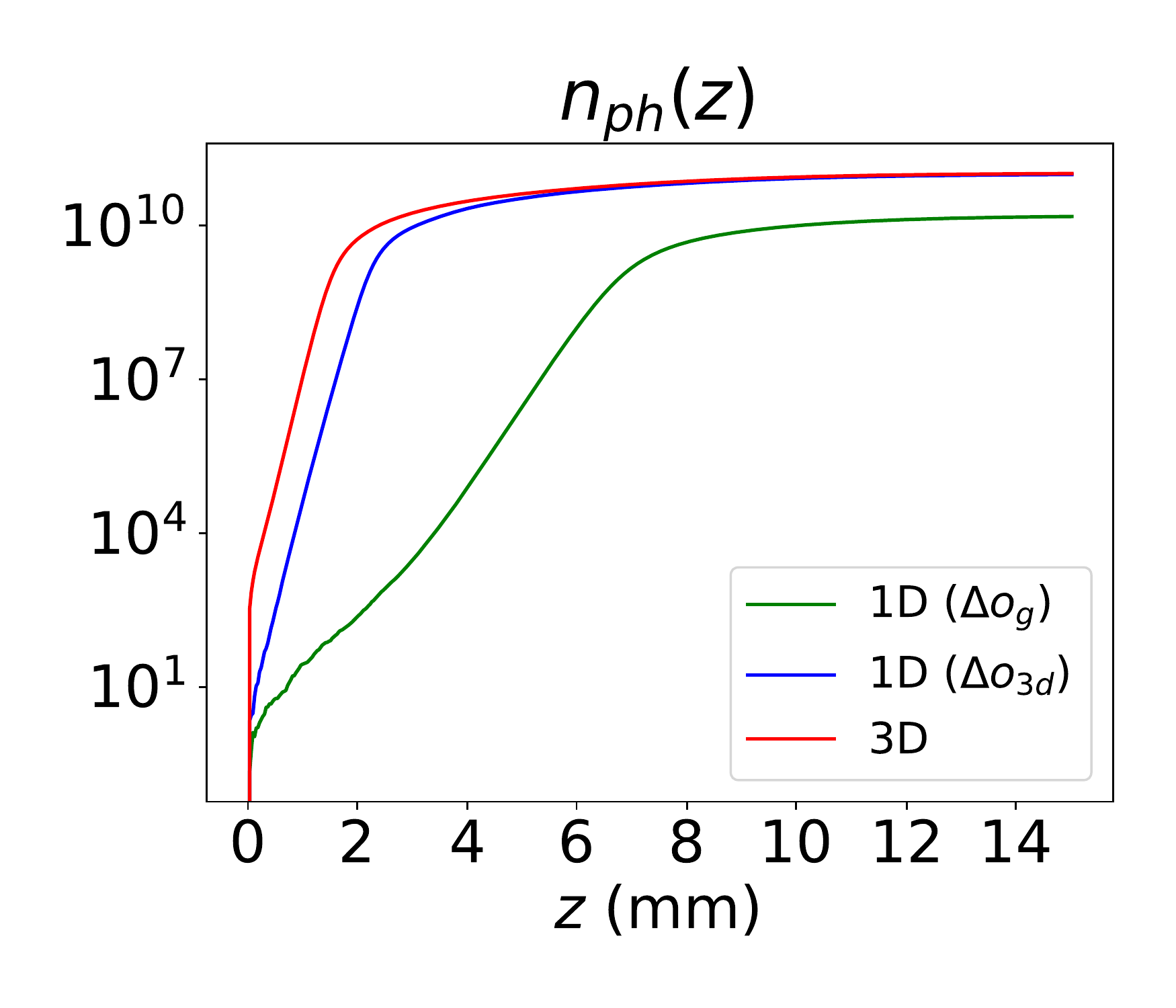}
         \caption{}
         \label{one1}
     \end{subfigure}\hfill
     \begin{subfigure}[t]{0.5\linewidth}
       \centering
\includegraphics[width=\textwidth]{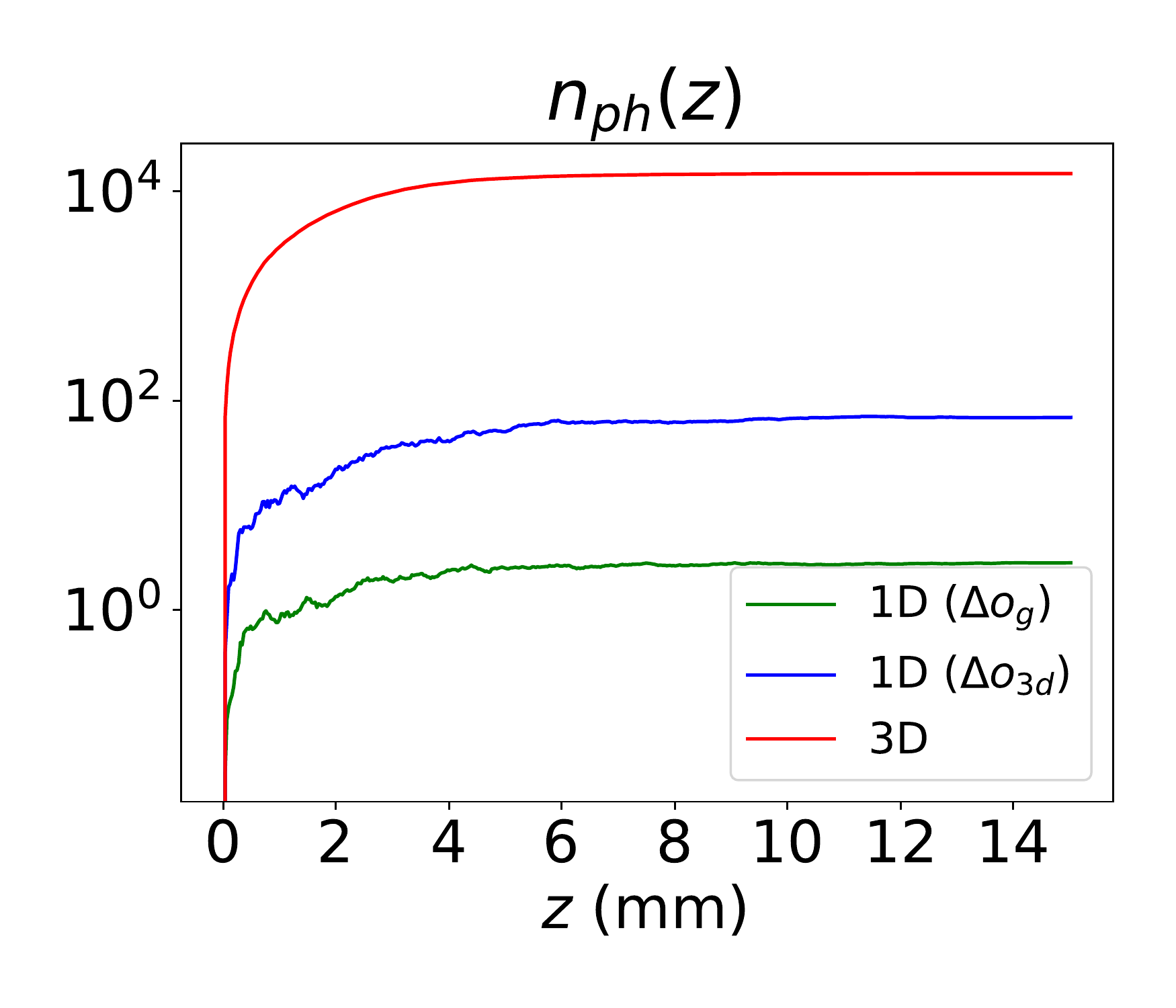}
         \caption{}
         \label{one2}
     \end{subfigure}\hfill
     \caption{Comparison between photon number predicted by modified 3D MBEs and modified 1D MBEs with \(\Delta o_g\) and \(\Delta o_{3d}\) for \(N_{ph,pump}\) of (a) \(2\times 10^{12}\) and (b) \(1 \times 10^{11}\).}
        \label{intensity}
    \end{figure}

Figure \ref{converge} shows the profile of normalized radiation intensity in \(\tau-z\) space computed with our modified MBE; panel (a) and (b) for 1D using \(\Delta o=\Delta o_g\) and \(\Delta o=\Delta o_{3d}\), which is about three times larger than \(\Delta o_g\), respectively, and panel (c) for 3D. The fluctuating feature prominent for \(z\ \lesssim 2~\mathrm{mm}\) for panel (a) is due to the random noise terms. After averaging over \(10^4\) samples, the profile becomes smooth (not shown) and becomes identical to that of obtained by the method of 1D two-point correlation function method shown in Fig. 3a in \cite{benediktovitch}. The profile obtained from our modified 3D MBEs, shown in panel (c), is quite different; The noisy region is confined to much smaller \(z\) and occupies a broader region in \(\tau\)   at a fixed \(z\).  The smoother 3D profile is partially due to the fact that the computation involves much more random number calls than 1D case. Note the profile in panel (b) is closer than panel (a) to that in the panel (c).  

The plot of the total number of photons, \(n(z)=\int d\tau I(z,\tau)\), is shown in Fig. \ref{intensity} (a), in which the red curve corresponds to Fig. \ref{converge} (c) while green curve  to Fig. \ref{converge} (a) and blue curve to Fig. \ref{converge} (b).  We see that the start-up photon number, which is due to spontaneous emission, is much larger for the 3D than the 1D case.  This is probably due to the fact that the angular spread of spontaneous emission is large. Indeed, Eq.~\eqref{paragreen} becomes large as \(z\) becomes small. The exponential growth is also much steeper for 3D than the 1D case--the gain is nonlinear. The blue curve is a much better approximation than the green curve to the red curve from 3D calculation.  We also conjecture that higher gain is responsible for \(\tau\)-broadening apparent in Fig.~\ref{converge} (c), since photons can be held longer against the decay/loss mechanism.  When the pump intensity is reduced by a factor of 20 the photon numbers at saturation is reduced by a factor of \(10^6\) as shown in Fig. \ref{intensity} (b). The differences between the red, blue, and green curves become more pronounced.

We presented a new Hamiltonian-based theory for X-ray ASE in paraxial approximation, extending the previous 1D analysis to include the diffraction effect in paraxial approximation and incorporating the spontaneous emission by adding a random noise term to MBEs.  The theory also includes the seed field and can serve as a practical tool to accurately model an X-ray ASE and XLO in either low or high-gain regime.

\section{acknowledgements}
We wish to thank  Kai Li  and  Linda Young at University of Chicago for introducing the physics of X-ray interaction and propagation through an atomic medium pumped by an XFEL, Andrei Benediktovitch and Nina Rohringer at DESY for extensive discussions on the 1D theory, and Alex Halavanau at SLAC for details of 1D simulation of XLO. This work is supported by the U.S. Department of Energy, Office of Science under Contract
No. DE-AC02-06CH11357.


\begin{thebibliography}{99}
\bibitem{footnote1}
To be consistent, the coefficient in front of the inhomogeneous term in \cite{benediktovitch} was reduced to one  half of that in Eq. (\ref{Eslow}).
\end{thebibliography}
\end{document}